\documentclass[aps,prl,twocolumn, superscriptaddress,showpacs]{revtex4-1}

\usepackage{graphicx}

\begin{document}

\title{Orbital dependent nucleonic pairing in the lightest known isotopes of tin}

\author{I.G.~Darby} 
\affiliation{Department of Physics and Astronomy, University of Tennessee, Knoxville, Tennessee 37996, USA} 
\affiliation{Instituut voor Kern- en Stralingsfysica, K.U.Leuven, Celestijnenlaan 200D, B-3001 Leuven, Belgium}

\author{R.K.~Grzywacz} 
\affiliation{Department of Physics and Astronomy, University of Tennessee, Knoxville, Tennessee 37996, USA} 
\affiliation{Physics Division, Oak Ridge National Laboratory, Oak Ridge, Tennessee 37831, USA}

\author{J.C.~Batchelder} 
\affiliation{UNIRIB, Oak Ridge Associated Universities, Oak Ridge, Tennessee 37831, USA}

\author{C.R.~Bingham} 
\affiliation{Department of Physics and Astronomy, University of Tennessee, Knoxville, Tennessee 37996, USA} 
\affiliation{Physics Division, Oak Ridge National Laboratory, Oak Ridge, Tennessee 37831, USA}

\author{L.~Cartegni} 
\affiliation{Department of Physics and Astronomy, University of Tennessee, Knoxville, Tennessee 37996, USA}

\author{C.J.~Gross} 
\affiliation{Physics Division, Oak Ridge National Laboratory, Oak Ridge, Tennessee 37831, USA}

\author{M.~Hjorth-Jensen} 
\affiliation{Department of Physics and Center of Mathematics for Applications, University of Oslo, N-0316 Oslo, Norway}

\author{D.T.~Joss} 
\affiliation{Oliver Lodge Laboratory, University of Liverpool,  Liverpool, L69 7ZE, UK}

\author{S.N.~Liddick} 

\affiliation{Department of Physics and Astronomy, University of Tennessee, Knoxville, Tennessee 37996, USA}

\author{W.~Nazarewicz} 
\affiliation{Department of Physics and Astronomy, University of Tennessee, Knoxville, Tennessee 37996, USA} 
\affiliation{Physics Division, Oak Ridge National Laboratory, Oak Ridge, Tennessee 37831, USA}
\affiliation{Institute of Theoretical Physics,  University of Warsaw, ul. Ho\.{z}a 69, PL-00681 Warsaw, Poland}

\author{S.~Padgett} 
\affiliation{Department of Physics and Astronomy, University of Tennessee, Knoxville, Tennessee 37996, USA}

\author{R.D.~Page} 
\affiliation{Oliver Lodge Laboratory, University of Liverpool,  Liverpool, L69 7ZE, UK}

\author{T.~Papenbrock} 
\affiliation{Department of Physics and Astronomy, University of Tennessee, Knoxville, Tennessee 37996, USA} 
\affiliation{Physics Division, Oak Ridge National Laboratory, Oak Ridge, Tennessee 37831, USA}

\author{M.M.~Rajabali} 
\affiliation{Department of Physics and Astronomy, University of Tennessee, Knoxville, Tennessee 37996, USA} 

\author{J.~Rotureau} 
\affiliation{Department of Physics and Astronomy, University of Tennessee, Knoxville, Tennessee 37996, USA} 

\author{K.P.~Rykaczewski}
\affiliation{Physics Division, Oak Ridge National Laboratory, Oak Ridge, Tennessee 37831, USA}

\date{\today}

\begin{abstract} 

By studying the  $^{109}$Xe$\rightarrow ^{105}$Te$\rightarrow ^{101}$Sn superallowed $\alpha$-decay chain, we observe low-lying states in $^{101}$Sn, the one-neutron system outside doubly magic $^{100}$Sn. We find that the spins of the ground state ($J=7/2$) and first excited state ($J=5/2$) in $^{101}$Sn are reversed with respect to the traditional level ordering postulated for $^{103}$Sn and the heavier tin isotopes. Through simple arguments and state-of-the-art shell model calculations we explain this unexpected switch in terms of a transition from the single-particle regime to the collective mode in which orbital-dependent pairing correlations, dominate.

\end{abstract}


\maketitle

Atomic nuclei exhibit a variety of motions, ranging from single-particle (s.p.) behaviour as dictated by the properties of very few nucleons, to collective phenomena which are emergent in character. In the nuclear realm, doubly magic nuclei with closed proton and neutron shells are of particular importance since they provide a shell-model framework~\cite{May72} through which we explain nuclear behaviour. Among magic species, the self-conjugated ($N=Z=50$) nucleus $^{100}$Sn is of special significance: revealed experimentally as a short-lived, 
neutron impoverished system predicted to be the endpoint of the most enhanced $\alpha$-decays known \cite{Lid06}. From a theoretical standpoint, $^{100}$Sn is the cornerstone of our understanding of nuclei within the entire $50 \leq N$, $Z\leq82$ region and a perfect laboratory for studying a variety of proton and neutron modes at the limits of particle stability.

The low-energy structure of semi-magic nuclei such as the tin isotopes is usually dominated by pairing correlations, or nucleonic superfluidity~\cite{Bri05}. In tin isotopes with an even number of neutrons, the $J^{\pi}$~=~0$^{\rm +}$ ground state (g.s.) can be viewed as a condensate of monopole (spin zero) Cooper pairs of valence neutrons (zero-quasiparticle state), while the lowest-lying states of neighbouring odd-mass isotopes can be viewed as one-quasiparticle states whose spins are determined by the angular momenta of the orbitals occupied by the unpaired neutron. Experimental data for the known semi-magic isotopic and isotonic chains indicate no exception to this rule, at least in the vicinity of shell closures.

The s.p. neutron states outside the $^{100}$Sn core are believed to be the closely-spaced $d_{5/2}$ and $g_{7/2}$ orbitals, consistent with the idea of pseudo-spin~\cite{Ari69,Hec69}. Thus, one expects the g.s.~spins for $^{\rm 101,103,105}$Sn to be identical and equal to $J^{\pi}$ = 5/2$^{+}$ or $J^{\pi}$ = 7/2$^{+}$, depending on whether $d_{5/2}$ or $g_{7/2}$ is the lowest-energy s.p. orbital. One of the main determinations of this work is the ordering of these crucial single-neutron levels in the nucleus $^{101}$Sn from $\alpha$-decay spectroscopy, which reveals that the usual shell-model extrapolation does not apply to the neutron-deficient tin isotopes. 

The identification of the two lowest states in $^{101}$Sn was carried out at the Holifield Radioactive Ion Beam Facility, Oak Ridge National Laboratory. The states in $^{101}$Sn were observed by studying the $\alpha$-decay chain $^{109}$Xe$\rightarrow$$^{105}$Te$\rightarrow$$^{101}$Sn \cite{Lid06} following heavy ion fusion evaporation reactions of $^{54}$Fe and $^{58}$Ni. Mass A~=~109 reaction products were resolved by atomic mass-to-ionic charge ratio and separated from un-reacted primary beam using the Recoil Mass Spectrometer (RMS)~\cite{Gro00}. The recoiling fusion evaporation residues were implanted at the RMS focal plane into a double-sided silicon strip detector (DSSD) where subsequent radioactive decays were observed. The experiments were instrumented with digital electronics~\cite{Grz03} capable of selectively capturing preamplifier double-pulse waveforms from very rapid sequential detector signals.
Two campaigns were performed: (a) low-rate high DSSD resolution and (b) high-rate $\alpha-\gamma$ coincidence. For (b) in addition to the DSSD, the ancillary detector suite CARDS~\cite{Kro02}, comprising four large volume HPGe Clover detectors for $\gamma$-ray detection, was placed around the DSSD chamber. Similar $\alpha$-decay statistics were obtained in both campaigns. The indirect method of producing $^{101}$Sn provided a mechanism for populating the $^{101}$Sn states and unambiguous isotope identification from the characteristic $\alpha$-decays of the parent and grandparent nuclei.

\begin{figure} \includegraphics[width=0.90\columnwidth]{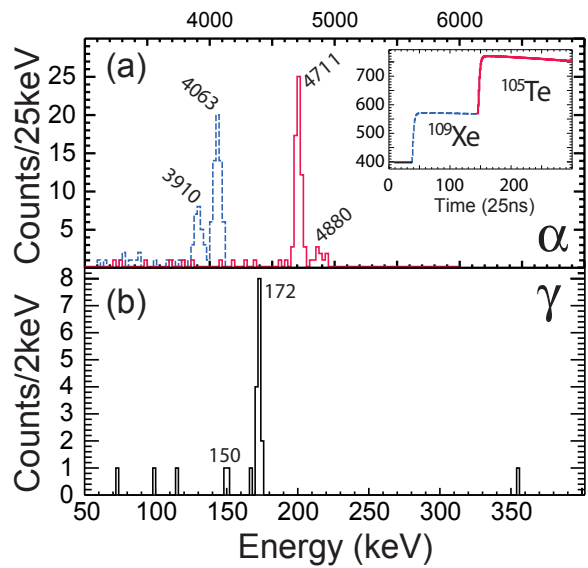} \caption{\label{fig1}(colour online) (a): High-resolution data $\alpha$-decay energy spectrum extracted from traces. The decays at 3910(10) keV and 4063(4) keV (blue dashed line) are assigned as $^{109}$Xe$\rightarrow$$^{105}$Te transitions. The decays at 4711(3) keV and 4880(20) keV (red solid line) are assigned as $^{105}$Te$\rightarrow$$^{101}$Sn. Inset: An example of a double-pulse preamplifier trace. The trace is highlighted showing the rises associated with  decays from $^{109}$Xe (first rise, blue dashed line) and $^{105}$Te (second, red solid line). (b): $\gamma$-ray spectrum in coincidence with double-$\alpha$ pulses. Two $\gamma$-ray lines at 150(3) keV and 172(2) keV are observed. } 
\end{figure} 
The $\alpha$-decay and $\gamma$-ray energy spectra associated with double-pulse events are shown in Fig.~\ref{fig1}. We observe four $\alpha$-decay transitions. Two $\alpha$-decays, 3910(10) keV and 4063(4) keV, extracted from the first pulse and assigned as the fine structure and g.s. $\alpha$-decay, respectively, from $^{109}$Xe yielding an excitation energy of 153(11) keV for the first excited state in $^{105}$Te, and a third decay, 4711(3) keV, extracted from the second pulse is assigned to $^{105}$Te. These decays are consistent with previously published values (3918(9), 4063(4) and 4703(5) keV~\cite{Lid06} and 4720(50) keV~\cite{Sew06}). Additionally, a fourth heretofore unknown $\alpha$-decay with energy 4880(20) keV has been extracted from the second pulse 
thus assigned to $^{105}$Te. Assuming 100\% $\alpha$-emission, the measured intensities of these decays yield the $\alpha$-decay branching ratios of 89(4)\% (4711 keV) and 11(4)\% (4880 keV). Assuming that these decays populate the ground and first excited states in $^{101}$Sn, the measured $\alpha$-decay energy difference yields a first excited state energy of 170(20) keV.

In coincidence with double-pulse events, we observed two $\gamma$-rays at 150(3) keV and 172(2) keV. The weaker 150 keV line is compatible with the depopulation of an excited state in $^{105}$Te, whose intensity, after efficiency corrections, corresponds to a 30\% $\alpha$-decay branch. The stronger 172 keV line is compatible with the depopulation of an excited state in $^{101}$Sn as indicated by the $\alpha$-decay energies and is consistent with the $\gamma$-ray of 172 keV, previously assigned as the de-excitation of the 1$^{\rm st}$-excited state in $^{101}$Sn~\cite{Sew07}. The characteristic double $\alpha$-decay pulse shapes provide a unique and clean coincidence requirement and allow us to assign this $\gamma$-ray unambiguously to $^{101}$Sn.

In order to account for the relatively large intensity of the observed 172 keV $\gamma$-ray transition, the excited state must be fed by the 4711 keV $\alpha$-decay. In light of this experimental finding, this $\alpha$-decay, previously assigned as the g.s.-to-g.s. transition~\cite{Lid06, Sew06} is now reinterpreted as the $\alpha$-decay from the ground state of $^{105}$Te to the {\it first excited state} in $^{101}$Sn. It is, therefore, the higher-energy 4880 keV decay that should be associated with the g.s.-to-g.s. transition.

\begin{figure} \includegraphics[width=0.85\columnwidth]{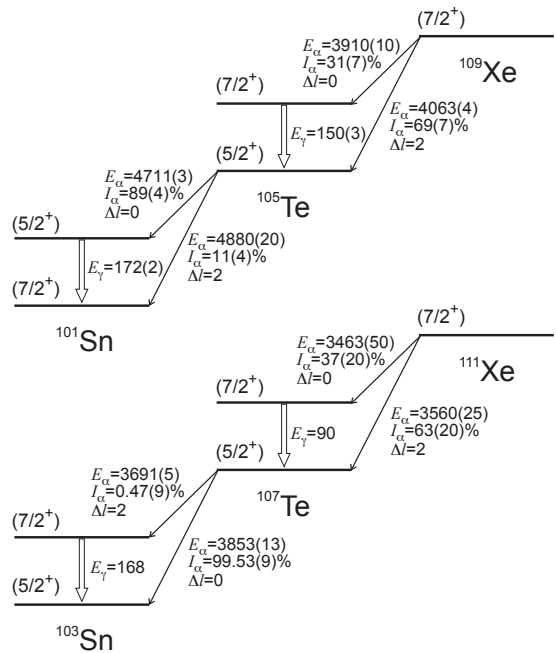} \caption{\label{fig2} Top: $\alpha$-decay chain $^{109}$Xe$\rightarrow$$^{105}$Te$\rightarrow$$^{101}$Sn proposed in this paper.  The large $\alpha$-decay branches, $I_\alpha$, for the 3910 keV and 4711 keV decays are interpreted as being due to the zero angular momentum ($\Delta l = 0$)  of the $\alpha$-decay transition. This compensates for the decreased $\alpha$-decay energy when compared to the 4063 keV and 4880 keV g.s.-to-g.s. $\Delta l=2$ transitions. Bottom: $\alpha$-decay chain $^{111}$Xe$\rightarrow$$^{107}$Te$\rightarrow$$^{103}$Sn from Refs.~\cite{Sch79,Sew02,Fah01}. All energies are in keV. }
\end{figure} 
The proposed decay scheme for the $^{109}$Xe $\alpha$-decay chain is shown in Fig.~\ref{fig2} (top). Interpreting this scheme within the standard model of $\alpha$-decay~\cite{Gam28} we are bound to conclude that the g.s. spins of $^{105}$Te ($N=53$) and $^{101}$Sn (N=51) must differ from each other, while the spin of the first excited state of $^{101}$Sn is equal to the spin of the $^{105}$Te g.s.

The neighbouring $\alpha$-decay chain $^{111}$Xe$\rightarrow ^{107}$Te $\rightarrow ^{103}$Sn \cite{Sch79,Sew02} is shown in Fig.~\ref{fig2} (bottom). In this case, only a very weak, ($I_{\alpha}$=0.5\%), $\Delta l=2$ fine structure $\alpha$-decay branch from the $^{107}$Te g.s. to the $^{103}$Sn excited state is observed, with the majority of the emission strength going via the higher-energy $\Delta l=0$, g.s.-to-g.s., transition. We note that while the first excited state in $^{103}$Sn at 168 keV~\cite{Sew02,Fah01} has an energy similar to that in $^{101}$Sn, the $\alpha$-decay pattern reveals that the g.s. spins of $^{107}$Te ($N=55$) and $^{103}$Sn ($N=53$) are the same, contrary to the situation in Fig.~\ref{fig2} (top).

The suggested level inversion observed in the $^{107}$Te$\rightarrow ^{103}$Sn and the $^{105}$Te$\rightarrow ^{101}$Sn decay chains is unexpected. In $^{109}$Sn, the g.s. spin has been measured as 5/2$^+$ with a first excited 7/2$^+$ state at 14 keV~\cite{Ebe87}. The energies of the first excited states in $^{103,105,107}$Sn have been measured and the g.s. spins tentatively assigned as 5/2$^+$, by considering systematic trends and/or theoretical predictions. Some information regarding the ground state in $^{101}$Sn has been previously obtained. Beta-delayed proton emission measurements~\cite{Kav07} resulted in a tentative assignment of a $J^{\pi}= 5/2^+$ g.s.. However, as noted Ref. \cite{Kav07}, the low statistics collected and the inherent ambiguities of the model used made this assignment inconclusive and the measurement is not incompatible with a $J^{\pi}= 7/2^{+}$ g.s.. A $J^{\pi}$=5/2$^+$ assignment has also been proposed~\cite{Sew07} following the observation of a solitary 172 keV $\gamma$-ray. They interpreted the non-observation of higher-energy $\gamma$-rays along with extrapolations from systematic trends in favour of $J^{\pi}= 5/2{^+}$.

In order to develop an understanding of the nature of the observed structural change between $^{103}$Sn and $^{101}$Sn, we present first a simple two-level model, based on the seniority scheme~\cite{Tal93}, which conveys much of the physics behind the experimental results. We consider a truncated configuration space, which consists only of the $g_{7/2}$ and $d_{5/2}$ orbitals. This approximation is justified for the light tin isotopes by the near degeneracy of these states and their large energy separation from the higher-lying orbitals. As the structure of semi magic nuclei is dominated by pairing, we further limit our discussion of $^{103}$Sn to the two main seniority-one neutron configurations, which are ($g_{7/2}$)$^{2}_{J=0}\otimes$$d_{5/2}$ and ($d_{5/2}$)$^{3}_{J=5/2}$ for states with $J^{\pi}= 5/2^+$, and ($d_{5/2}$)$^{2}_{J=0}\otimes$$g_{7/2}$ and ($g_{7/2}$)$^{3}_{J=7/2}$ for $J^{\pi} = 7/2^+$ states, respectively. The structure of the collective $J^{\pi} =0^+$ Cooper pair in the ground state of $^{102}$Sn is determined by the competition between the spacing $\Delta\varepsilon_{\rm sp}\approx$172 keV of the $g_{7/2}$ and $d_{5/2}$ levels and the pairing energies of the neutron pairs. The latter are primarily determined by the two diagonal two-body matrix elements (TBME) $V^{\rm pair}(l_j)=\langle(l_j)^{2}_{J=0}|V|(l_j)^{2}_{J=0}\rangle$  and one off-diagonal TBME $V^{\rm pair}(g_{7/2},d_{5/2})$ representing the scattering of $J^{\pi} =0^+$ pairs between $g_{7/2}$  and $d_{5/2}$ shells.
These matrix elements are computed~\cite{HjJ95} from the nucleon-nucleon potentials N3LO~\cite{Ent03} and
AV18~\cite{Wir95}. For N3LO, we find that  $V^{\rm pair}(g_{7/2})=1.40$\,MeV is significantly larger than $V^{\rm~pair}(d_{5/2})=0.84$\,MeV, and their difference $\Delta V^{\rm pair} =0.56$\,MeV is considerably larger than $\Delta\varepsilon_{\rm sp}$. 
(Similar results are obtained for AV18.)
In addition, since $\Delta V^{\rm pair} \approx | V^{\rm pair}(g_{7/2},d_{5/2})|$, substantial mixing between the $(g_{7/2})^{2}_{J=0}$ and $(d_{5/2})^{2}_{J=0}$ occurs.

The resulting structure of the Cooper pair in $^{102}$Sn is dominated by the ($g_{7/2}$)$^{2}_{J=0}$ component (about 70\%). When coupling the odd neutron to this collective pair to form the low-lying states in $^{103}$Sn, the Pauli blocking kicks in. In the 5/2$^+$ state, the weight of the ($g_{7/2})^{2}_{J=0}\otimes d_{5/2}$ configuration (83\%) becomes {\it enhanced} with respect to the ($g_{7/2})^{2}_{J=0}$ component in $^{102}$Sn, and due to the strong $g_{7/2}$ pairing, produces significant additional binding. Likewise, for the 7/2$^+$ state, the weight of the $(g_{7/2})^{3}_{J=7/2}$ configuration (33\%) becomes significantly {\it reduced}, thus lowering the binding. Consequently, it is the strong pairing energy in $(g_{7/2})^{2}_{J=0}$ and Pauli blocking that sets the $J^{\pi}=5/2^+$ g.s. spin in $^{103}$Sn. It appears that the region of $^{100}$Sn is quite unique in exhibiting such a behaviour. A large difference in pairing matrix elements of the pseudo-spin partners, going well beyond the usual $(2j+1)$ scaling \cite{Tal93} is unexpected from commonly-used phenomenological shell-model interactions.

We substantiate the insights from the two-level model by state-of-the-art configuration interaction (CI) calculations following Ref.~\cite{HjJ95}. These shell-model results are intended to lend further support to the experimental interpretation. In the first calculation variant (V1), we assumed a $^{100}$Sn core with valence nucleons in $d_{5/2}$, $g_{7/2}$, $d_{3/2}$, $s_{1/2}$ and $h_{11/2}$ shells. In the second variant (V2), we took an $^{88}$Sr core ($N=50$) with valence protons in $p_{1/2}$, $g_{9/2}$ shells and valence neutrons in $d_{5/2}$, $g_{7/2}$, $d_{3/2}$, $s_{1/2}$ and $h_{11/2}$  shells.  In V1, the residual interaction was based on  AV18 or N3LO nucleon-nucleon potentials and for V2 it was derived from the CD-Bonn potential~\cite{Mac01}. All these interaction models contain no free parameters and are expected to describe accurately the structure of light systems~\cite{Nav09} and nuclei with sufficiently small numbers of valence nucleons. Results of calculations around $^{100}$Sn using V2, relevant in the context of this study, have been previously discussed~\cite{Lip02,Eks09}. Overall, V2 gives a very good agreement for $A \sim 100$ nuclei.

\begin{figure} \includegraphics[width=0.85\columnwidth]{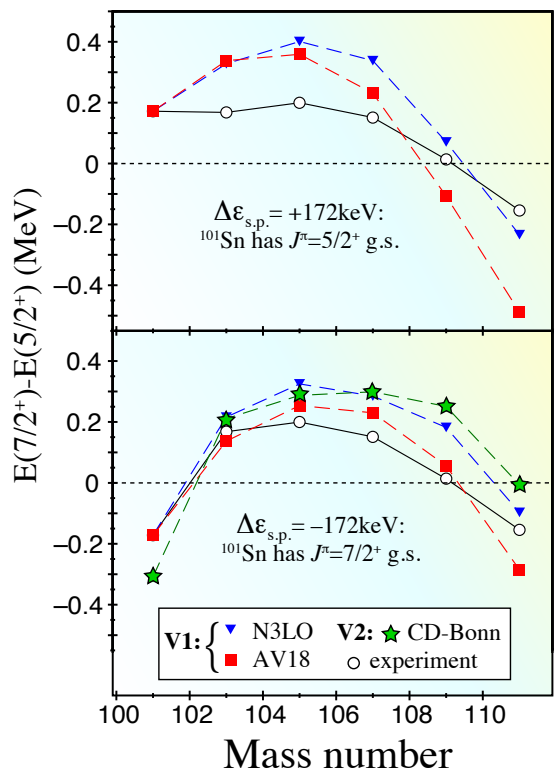} \caption{\label{fig3} (colour online) 
Results of our shell model calculations in variants V1 and V2 for the splitting of the $7/2^+ - 5/2^+$ states in the 
neutron-deficient, odd-mass tin isotopes. In V1, the upper (lower) panel assumes the $d_{5/2}$ single-neutron level above (below)  the $g_{7/2}$ level. Experimental values (circles) are taken from this work and from Ref.~\cite{Fah01} and references quoted therein.  
}
\end{figure} 

Figure~\ref{fig3} compares results of V1 and V2 with experiment. In V1, the splitting $|\Delta\varepsilon_{\rm sp}|$ has been set to the experimental value of 172 keV. Calculations are performed for both possible orderings of the levels, thus allowing for either 7/2$^+$ or 5/2$^+$ ground state in $^{101}$Sn. However, {\it regardless of this ordering}, the ground state of the heavier Sn isotopes, in particular $^{103}$Sn, always turns out to be 5/2$^+$. Assuming a 5/2$^+$ ground state in $^{101}$Sn, our calculations overestimate the location of the 1$^{\rm st}$-excited state in the heavier Sn isotopes by about 200 keV. Assuming a 7/2$^+$ ground state, on the other hand, we find excellent agreement between theory and experiment. In V2, based on experimental s.p. energies of $^{89}$Sr (neutrons) and $^{89}$Y (protons), the ground state of $^{101}$Sn is predicted to be 7/2$^+$, and this is consistent with the previous study~\cite{Gra06}. Both V1 and V2 reproduce well the experimental parabolic trend of the excited states, including the crossing between 5/2$^+$ and 7/2$^+$ ground states between $^{109}$Sn and $^{111}$Sn, as well as the energy of the lowest 2$^+$  seniority-two state in the even-mass tin isotopes (not shown in Fig.~\ref{fig3}).

It has been suggested in Ref.~\cite{Sew07} that good agreement can also be
achieved assuming a 5/2$^+$ g.s. by reducing  $V^{\rm pair}(g_{7/2})$ by about
30\%. Such an adjustment is not consistent with the microscopically derived
residual interactions. 
It cannot be excluded that the 5/2$^+$ and 7/2$^+$ level inversion occurs between $^{109}$Xe and $^{105}$Te. However, we do not
find evidence to support such a scenario. In Fig.~\ref{fig2}, we see that
the lower-energy $\alpha$-decays from $^{109}$Xe and $^{111}$Xe populate the
excited states of $^{105}$Te and $^{107}$Te, respectively, presenting no
indication of a change in structure in tellurium. This is supported by the CI
calculations for the tellurium isotopes: in V1+AV18 we predict 5/2$^+$ for the ground
states of $^{105}$Te and $^{107}$Te, regardless of the sign of
$\Delta\varepsilon_{\rm sp}$ or even with 30\% reduced 
$V^{\rm pair}(g_{7/2})$~\cite{Sew07}.

In conclusion, our $\alpha$-decay studies give strong experimental evidence for a 7/2$^+$ g.s and 5/2$^+$ first excited state in $^{101}$Sn. This is contrary to what has been previously postulated, based on extrapolation from the heavier tin isotopes. Shell model calculations with realistic interactions, both in a two-level space and in a large configuration space, strongly support our new interpretation. The inversion of the g.s. spins between $^{103}$Sn and $^{101}$Sn is due to the unusually strong pairing interaction between the $g_{7/2}$ neutrons and unusually small energy splitting between the 7/2$^+$ and 5/2$^+$ states in $^{101}$Sn. The strong pairing in $g_{7/2}$  above $^{100}$Sn can be related to significant contributions from the two-body tensor force,
which is expected~\cite{Ots10} to  produce  7/2$^+$--5/2$^+$ level inversion in $^{101}$Sn, in good agreement with our analysis.
The region of proton-rich nuclei above $^{100}$Sn seems to be quite unique in
exhibiting such unusual behaviour.
Interestingly, another doubly magic isotope of tin, the neutron rich  $^{132}$Sn, is a well-behaved shell-model system because of the large energy  splitting between single particle orbitals \cite{Jon10}.  

This research is sponsored by the Office of Science, U.S. Department of Energy under contracts DE-FG02-96ER40983 (UT), DE-AC05-00OR22725 (ORNL), DE-AC05-060R23100 (ORAU) and DE-FC03-03NA00143 (NNSA); and by the UK Science and Technology Facilities Council. 


\bibliography{sn101}

\begin{thebibliography}{27}%
\makeatletter
\providecommand \@ifxundefined [1]{%
 \@ifx{#1\undefined}
}%
\providecommand \@ifnum [1]{%
 \ifnum #1\expandafter \@firstoftwo
 \else \expandafter \@secondoftwo
 \fi
}%
\providecommand \@ifx [1]{%
 \ifx #1\expandafter \@firstoftwo
 \else \expandafter \@secondoftwo
 \fi
}%
\providecommand \natexlab [1]{#1}%
\providecommand \enquote  [1]{``#1''}%
\providecommand \bibnamefont  [1]{#1}%
\providecommand \bibfnamefont [1]{#1}%
\providecommand \citenamefont [1]{#1}%
\providecommand \href@noop [0]{\@secondoftwo}%
\providecommand \href [0]{\begingroup \@sanitize@url \@href}%
\providecommand \@href[1]{\@@startlink{#1}\@@href}%
\providecommand \@@href[1]{\endgroup#1\@@endlink}%
\providecommand \@sanitize@url [0]{\catcode `\\12\catcode `\$12\catcode
  `\&12\catcode `\#12\catcode `\^12\catcode `\_12\catcode `\%12\relax}%
\providecommand \@@startlink[1]{}%
\providecommand \@@endlink[0]{}%
\providecommand \url  [0]{\begingroup\@sanitize@url \@url }%
\providecommand \@url [1]{\endgroup\@href {#1}{\urlprefix }}%
\providecommand \urlprefix  [0]{URL }%
\providecommand \Eprint [0]{\href }%
\@ifxundefined \urlstyle {%
  \providecommand \doi  [0]{\begingroup \@sanitize@url \@doi}%
  \providecommand \@doi [1]{\endgroup \@@startlink {\doibase
  #1}doi:\discretionary {}{}{}#1\@@endlink }%
}{%
  \providecommand \doi  [0]{doi:\discretionary{}{}{}\begingroup
  \urlstyle{rm}\Url }%
}%
\providecommand \doibase [0]{http://dx.doi.org/}%
\providecommand \Doi [0]{\begingroup \@sanitize@url \@Doi }%
\providecommand \@Doi  [1]{\endgroup\@@startlink{\doibase#1}\@@Doi}%
\providecommand \@@Doi [1]{#1\@@endlink}%
\providecommand \selectlanguage [0]{\@gobble}%
\providecommand \bibinfo  [0]{\@secondoftwo}%
\providecommand \bibfield  [0]{\@secondoftwo}%
\providecommand \translation [1]{[#1]}%
\providecommand \BibitemOpen [0]{}%
\providecommand \bibitemStop [0]{}%
\providecommand \bibitemNoStop [0]{.\EOS\space}%
\providecommand \EOS [0]{\spacefactor3000\relax}%
\providecommand \BibitemShut  [1]{\csname bibitem#1\endcsname}%
\bibitem [{\citenamefont {Mayer}\ and\ \citenamefont {Jensen}(1972)}]{May72}%
  \BibitemOpen
  \bibfield  {author} {\bibinfo {author} {\bibfnamefont {M.~G.}\ \bibnamefont
  {Mayer}}\ and\ \bibinfo {author} {\bibfnamefont {J.~H.~D.}\ \bibnamefont
  {Jensen}},\ }\href@noop {} {}\bibinfo {howpublished} {Nobel Lectures, Physics
  1963-1970, Elsevier Amsterdam} (\bibinfo {year} {1972})\BibitemShut {NoStop}%
\bibitem [{\citenamefont {Liddick}\ \emph {et~al.}(2006)\citenamefont {Liddick}
  \emph {et~al.}}]{Lid06}%
  \BibitemOpen
  \bibfield  {author} {\bibinfo {author} {\bibfnamefont {S.~N.}\ \bibnamefont
  {Liddick}} \emph {et~al.},\ }\href@noop {} {\bibfield  {journal} {\bibinfo
  {journal} {Phys. Rev. Lett.},\ }\textbf {\bibinfo {volume} {97}},\ \bibinfo
  {pages} {082501} (\bibinfo {year} {2006})}\BibitemShut {NoStop}%
\bibitem [{\citenamefont {Brink}\ and\ \citenamefont {Broglia}(2005)}]{Bri05}%
  \BibitemOpen
  \bibfield  {author} {\bibinfo {author} {\bibfnamefont {D.~M.}\ \bibnamefont
  {Brink}}\ and\ \bibinfo {author} {\bibfnamefont {R.~A.}\ \bibnamefont
  {Broglia}},\ }\href@noop {} {\emph {\bibinfo {title} {Nuclear Superfluidity:
  pairing in finite systems}}}\ (\bibinfo  {publisher} {Cambridge Univ.
  Press},\ \bibinfo {year} {2005})\BibitemShut {NoStop}%
\bibitem [{\citenamefont {Arima}\ \emph {et~al.}(1969)\citenamefont {Arima},
  \citenamefont {Harvey},\ and\ \citenamefont {Shimizu}}]{Ari69}%
  \BibitemOpen
  \bibfield  {author} {\bibinfo {author} {\bibfnamefont {A.}~\bibnamefont
  {Arima}}, \bibinfo {author} {\bibfnamefont {M.}~\bibnamefont {Harvey}}, \
  and\ \bibinfo {author} {\bibfnamefont {K.}~\bibnamefont {Shimizu}},\
  }\href@noop {} {\bibfield  {journal} {\bibinfo  {journal} {Phys. Lett. B},\
  }\textbf {\bibinfo {volume} {30}},\ \bibinfo {pages} {517} (\bibinfo {year}
  {1969})}\BibitemShut {NoStop}%
\bibitem [{\citenamefont {Hecht}\ and\ \citenamefont {Adler}(1969)}]{Hec69}%
  \BibitemOpen
  \bibfield  {author} {\bibinfo {author} {\bibfnamefont {K.~T.}\ \bibnamefont
  {Hecht}}\ and\ \bibinfo {author} {\bibfnamefont {A.}~\bibnamefont {Adler}},\
  }\href@noop {} {\bibfield  {journal} {\bibinfo  {journal} {Nucl. Phys. A},\
  }\textbf {\bibinfo {volume} {137}},\ \bibinfo {pages} {129} (\bibinfo {year}
  {1969})}\BibitemShut {NoStop}%
\bibitem [{\citenamefont {Gross}\ \emph {et~al.}(2000)\citenamefont {Gross}
  \emph {et~al.}}]{Gro00}%
  \BibitemOpen
  \bibfield  {author} {\bibinfo {author} {\bibfnamefont {C.~J.}\ \bibnamefont
  {Gross}} \emph {et~al.},\ }\href@noop {} {\bibfield  {journal} {\bibinfo
  {journal} {Nucl. Instr. Meth. A},\ }\textbf {\bibinfo {volume} {450}},\
  \bibinfo {pages} {12} (\bibinfo {year} {2000})}\BibitemShut {NoStop}%
\bibitem [{\citenamefont {Grzywacz}(2003)}]{Grz03}%
  \BibitemOpen
  \bibfield  {author} {\bibinfo {author} {\bibfnamefont {R.}~\bibnamefont
  {Grzywacz}},\ }\href@noop {} {\bibfield  {journal} {\bibinfo  {journal}
  {Nucl. Instr. Meth. B},\ }\textbf {\bibinfo {volume} {204}},\ \bibinfo
  {pages} {649} (\bibinfo {year} {2003})}\BibitemShut {NoStop}%
\bibitem [{\citenamefont {Krolas}\ \emph {et~al.}(2002)\citenamefont {Krolas}
  \emph {et~al.}}]{Kro02}%
  \BibitemOpen
  \bibfield  {author} {\bibinfo {author} {\bibfnamefont {W.}~\bibnamefont
  {Krolas}} \emph {et~al.},\ }\href@noop {} {\bibfield  {journal} {\bibinfo
  {journal} {Phys. Rev. C},\ }\textbf {\bibinfo {volume} {65}},\ \bibinfo
  {pages} {031303} (\bibinfo {year} {2002})}\BibitemShut {NoStop}%
\bibitem [{\citenamefont {Seweryniak}\ \emph {et~al.}(2006)\citenamefont
  {Seweryniak} \emph {et~al.}}]{Sew06}%
  \BibitemOpen
  \bibfield  {author} {\bibinfo {author} {\bibfnamefont {D.}~\bibnamefont
  {Seweryniak}} \emph {et~al.},\ }\href@noop {} {\bibfield  {journal} {\bibinfo
   {journal} {Phys. Rev. C},\ }\textbf {\bibinfo {volume} {73}},\ \bibinfo
  {pages} {061301(R)} (\bibinfo {year} {2006})}\BibitemShut {NoStop}%
\bibitem [{\citenamefont {Seweryniak}\ \emph {et~al.}(2007)\citenamefont
  {Seweryniak} \emph {et~al.}}]{Sew07}%
  \BibitemOpen
  \bibfield  {author} {\bibinfo {author} {\bibfnamefont {D.}~\bibnamefont
  {Seweryniak}} \emph {et~al.},\ }\href@noop {} {\bibfield  {journal} {\bibinfo
   {journal} {Phys. Rev. Lett.},\ }\textbf {\bibinfo {volume} {99}},\ \bibinfo
  {pages} {022504} (\bibinfo {year} {2007})}\BibitemShut {NoStop}%
\bibitem [{\citenamefont {Schardt}\ \emph {et~al.}(1979)\citenamefont {Schardt}
  \emph {et~al.}}]{Sch79}%
  \BibitemOpen
  \bibfield  {author} {\bibinfo {author} {\bibfnamefont {D.}~\bibnamefont
  {Schardt}} \emph {et~al.},\ }\href@noop {} {\bibfield  {journal} {\bibinfo
  {journal} {Nucl. Phys. A},\ }\textbf {\bibinfo {volume} {326}},\ \bibinfo
  {pages} {65} (\bibinfo {year} {1979})}\BibitemShut {NoStop}%
\bibitem [{\citenamefont {Seweryniak}\ \emph {et~al.}(2002)\citenamefont
  {Seweryniak} \emph {et~al.}}]{Sew02}%
  \BibitemOpen
  \bibfield  {author} {\bibinfo {author} {\bibfnamefont {D.}~\bibnamefont
  {Seweryniak}} \emph {et~al.},\ }\href@noop {} {\bibfield  {journal} {\bibinfo
   {journal} {Phys. Rev. C},\ }\textbf {\bibinfo {volume} {66}},\ \bibinfo
  {pages} {051307(R)} (\bibinfo {year} {2002})}\BibitemShut {NoStop}%
\bibitem [{\citenamefont {Fahlander}\ \emph {et~al.}(2001)\citenamefont
  {Fahlander} \emph {et~al.}}]{Fah01}%
  \BibitemOpen
  \bibfield  {author} {\bibinfo {author} {\bibfnamefont {C.}~\bibnamefont
  {Fahlander}} \emph {et~al.},\ }\href@noop {} {\bibfield  {journal} {\bibinfo
  {journal} {Phys. Rev. C},\ }\textbf {\bibinfo {volume} {63}},\ \bibinfo
  {pages} {021307(R)} (\bibinfo {year} {2001})}\BibitemShut {NoStop}%
\bibitem [{\citenamefont {Gamow}(1928)}]{Gam28}%
  \BibitemOpen
  \bibfield  {author} {\bibinfo {author} {\bibfnamefont {G.}~\bibnamefont
  {Gamow}},\ }\href@noop {} {\bibfield  {journal} {\bibinfo  {journal} {Z.
  Phys.},\ }\textbf {\bibinfo {volume} {51}},\ \bibinfo {pages} {204} (\bibinfo
  {year} {1928})}\BibitemShut {NoStop}%
\bibitem [{\citenamefont {Eberz}\ \emph {et~al.}(1987)\citenamefont {Eberz}
  \emph {et~al.}}]{Ebe87}%
  \BibitemOpen
  \bibfield  {author} {\bibinfo {author} {\bibfnamefont {J.}~\bibnamefont
  {Eberz}} \emph {et~al.},\ }\href@noop {} {\bibfield  {journal} {\bibinfo
  {journal} {Zeits. Phys. A},\ }\textbf {\bibinfo {volume} {326}},\ \bibinfo
  {pages} {121} (\bibinfo {year} {1987})}\BibitemShut {NoStop}%
\bibitem [{\citenamefont {Kavatsyuk}\ \emph {et~al.}(2007)\citenamefont
  {Kavatsyuk} \emph {et~al.}}]{Kav07}%
  \BibitemOpen
  \bibfield  {author} {\bibinfo {author} {\bibfnamefont {O.}~\bibnamefont
  {Kavatsyuk}} \emph {et~al.},\ }\href@noop {} {\bibfield  {journal} {\bibinfo
  {journal} {Eur. Phys. J. A},\ }\textbf {\bibinfo {volume} {31}},\ \bibinfo
  {pages} {319} (\bibinfo {year} {2007})}\BibitemShut {NoStop}%
\bibitem [{\citenamefont {Talmi}(1993)}]{Tal93}%
  \BibitemOpen
  \bibfield  {author} {\bibinfo {author} {\bibfnamefont {I.}~\bibnamefont
  {Talmi}},\ }\href@noop {} {\emph {\bibinfo {title} {Simple models of complex
  nuclei.}}}\ (\bibinfo  {publisher} {Taylor \& Francis Ltd, London},\ \bibinfo
  {year} {1993})\BibitemShut {NoStop}%
\bibitem [{\citenamefont {Hjorth-Jensen}\ \emph {et~al.}(1995)\citenamefont
  {Hjorth-Jensen}, \citenamefont {Kuo},\ and\ \citenamefont {Osnes}}]{HjJ95}%
  \BibitemOpen
  \bibfield  {author} {\bibinfo {author} {\bibfnamefont {M.}~\bibnamefont
  {Hjorth-Jensen}}, \bibinfo {author} {\bibfnamefont {T.~T.~S.}\ \bibnamefont
  {Kuo}}, \ and\ \bibinfo {author} {\bibfnamefont {E.}~\bibnamefont {Osnes}},\
  }\href@noop {} {\bibfield  {journal} {\bibinfo  {journal} {Phys. Rep.},\
  }\textbf {\bibinfo {volume} {261}},\ \bibinfo {pages} {125} (\bibinfo {year}
  {1995})}\BibitemShut {NoStop}%
\bibitem [{\citenamefont {Entem}\ and\ \citenamefont
  {Machleidt}(2003)}]{Ent03}%
  \BibitemOpen
  \bibfield  {author} {\bibinfo {author} {\bibfnamefont {D.~R.}\ \bibnamefont
  {Entem}}\ and\ \bibinfo {author} {\bibfnamefont {R.}~\bibnamefont
  {Machleidt}},\ }\href@noop {} {\bibfield  {journal} {\bibinfo  {journal}
  {Phys. Rev. C},\ }\textbf {\bibinfo {volume} {68}},\ \bibinfo {pages}
  {041001(R)} (\bibinfo {year} {2003})}\BibitemShut {NoStop}%
\bibitem [{\citenamefont {Wiringa}\ \emph {et~al.}(1995)\citenamefont
  {Wiringa}, \citenamefont {Stoks},\ and\ \citenamefont {Schiavilla}}]{Wir95}%
  \BibitemOpen
  \bibfield  {author} {\bibinfo {author} {\bibfnamefont {R.~B.}\ \bibnamefont
  {Wiringa}}, \bibinfo {author} {\bibfnamefont {V.~G.~J.}\ \bibnamefont
  {Stoks}}, \ and\ \bibinfo {author} {\bibfnamefont {R.}~\bibnamefont
  {Schiavilla}},\ }\href@noop {} {\bibfield  {journal} {\bibinfo  {journal}
  {Phys. Rev. C},\ }\textbf {\bibinfo {volume} {51}},\ \bibinfo {pages} {38}
  (\bibinfo {year} {1995})}\BibitemShut {NoStop}%
\bibitem [{\citenamefont {Machleidt}(2001)}]{Mac01}%
  \BibitemOpen
  \bibfield  {author} {\bibinfo {author} {\bibfnamefont {R.}~\bibnamefont
  {Machleidt}},\ }\href@noop {} {\bibfield  {journal} {\bibinfo  {journal}
  {Phys. Rev. C},\ }\textbf {\bibinfo {volume} {63}},\ \bibinfo {pages}
  {024001} (\bibinfo {year} {2001})}\BibitemShut {NoStop}%
\bibitem [{\citenamefont {Navratil}\ \emph {et~al.}(2009)\citenamefont
  {Navratil} \emph {et~al.}}]{Nav09}%
  \BibitemOpen
  \bibfield  {author} {\bibinfo {author} {\bibfnamefont {P.}~\bibnamefont
  {Navratil}} \emph {et~al.},\ }\href@noop {} {\bibfield  {journal} {\bibinfo
  {journal} {J. Phys. G},\ }\textbf {\bibinfo {volume} {36}},\ \bibinfo {pages}
  {083101} (\bibinfo {year} {2009})}\BibitemShut {NoStop}%
\bibitem [{\citenamefont {Lipoglavsek}\ \emph {et~al.}(2002)\citenamefont
  {Lipoglavsek} \emph {et~al.}}]{Lip02}%
  \BibitemOpen
  \bibfield  {author} {\bibinfo {author} {\bibfnamefont {M.}~\bibnamefont
  {Lipoglavsek}} \emph {et~al.},\ }\href@noop {} {\bibfield  {journal}
  {\bibinfo  {journal} {Phys. Rev. C},\ }\textbf {\bibinfo {volume} {66}},\
  \bibinfo {pages} {011302(R)} (\bibinfo {year} {2002})}\BibitemShut {NoStop}%
\bibitem [{\citenamefont {Ekstrom}\ \emph {et~al.}(2009)\citenamefont {Ekstrom}
  \emph {et~al.}}]{Eks09}%
  \BibitemOpen
  \bibfield  {author} {\bibinfo {author} {\bibfnamefont {A.}~\bibnamefont
  {Ekstrom}} \emph {et~al.},\ }\href@noop {} {\bibfield  {journal} {\bibinfo
  {journal} {Phys. Rev. C},\ }\textbf {\bibinfo {volume} {80}},\ \bibinfo
  {pages} {054302} (\bibinfo {year} {2009})}\BibitemShut {NoStop}%
\bibitem [{\citenamefont {Grawe}\ \emph {et~al.}(2006)\citenamefont {Grawe}
  \emph {et~al.}}]{Gra06}%
  \BibitemOpen
  \bibfield  {author} {\bibinfo {author} {\bibfnamefont {H.}~\bibnamefont
  {Grawe}} \emph {et~al.},\ }\href@noop {} {\bibfield  {journal} {\bibinfo
  {journal} {Eur. Phys. J. A},\ }\textbf {\bibinfo {volume} {27 s01}},\
  \bibinfo {pages} {257} (\bibinfo {year} {2006})}\BibitemShut {NoStop}%
\bibitem [{\citenamefont {Otsuka}\ \emph {et~al.}(2010)\citenamefont {Otsuka}
  \emph {et~al.}}]{Ots10}%
  \BibitemOpen
  \bibfield  {author} {\bibinfo {author} {\bibfnamefont {T.}~\bibnamefont
  {Otsuka}} \emph {et~al.},\ }\href@noop {} {\bibfield  {journal} {\bibinfo
  {journal} {Phys. Rev. Lett.},\ }\textbf {\bibinfo {volume} {104}},\ \bibinfo
  {pages} {012501} (\bibinfo {year} {2010})}\BibitemShut {NoStop}%
\bibitem [{\citenamefont {Jones}\ \emph {et~al.}(2010)\citenamefont {Jones}
  \emph {et~al.}}]{Jon10}%
  \BibitemOpen
  \bibfield  {author} {\bibinfo {author} {\bibfnamefont {K.~L.}\ \bibnamefont
  {Jones}} \emph {et~al.},\ }\href@noop {} {\bibfield  {journal} {\bibinfo
  {journal} {Nature},\ }\textbf {\bibinfo {volume} {465}},\ \bibinfo {pages}
  {454} (\bibinfo {year} {2010})}\BibitemShut {NoStop}%
\end{thebibliography}%

\end{document}